\documentclass[aps,prl,twocolumn,superscriptaddress,showpacs]{revtex4-1}
\usepackage{graphicx,amsmath,amssymb,dcolumn,bm,array}
\newcommand{\iso}[1]{\textsuperscript{#1}}
\newcommand{\sceptar}[0]{{\scshape Sceptar}}
\newcommand{\isac}[0]{{\scshape Isac}}
\newcommand{\triumf}[0]{{\scshape Triumf}}

\newcolumntype{.}{D{.}{.}{-1}}
\begin{document}
\title{\boldmath High-Precision Measurement of the $^{19}$Ne Half-Life and Implications for Right-Handed Weak Currents
} 
\author{S.~Triambak} 
\email[]{smarajit@gmail.com}
\affiliation{TRIUMF, 4004 Wesbrook Mall, Vancouver, British Columbia V6T 2A3, Canada} 
\affiliation{Department of Physics \& Astrophysics, University of Delhi, Delhi 100 007, India}
\author{P.~Finlay} 
\affiliation{Department of Physics, University of Guelph, Guelph, Ontario N1G 2W1, Canada}
\author{C.\,S.~Sumithrarachchi} 
\altaffiliation[Present address: ]{National Superconducting Cyclotron Laboratory, Michigan State University, East Lansing, Michigan 48824, USA.}
\affiliation{Department of Physics, University of Guelph, Guelph, Ontario N1G 2W1, Canada}
\author{G.~Hackman} 
\affiliation{TRIUMF, 4004 Wesbrook Mall, Vancouver, British Columbia V6T 2A3, Canada}
\author{G.\,C.~Ball} 
\affiliation{TRIUMF, 4004 Wesbrook Mall, Vancouver, British Columbia V6T 2A3, Canada}
\author{P.\,E.~Garrett}
\affiliation{Department of Physics, University of Guelph, Guelph, Ontario N1G 2W1, Canada}
\author{C.\,E.~Svensson} 
\affiliation{Department of Physics, University of Guelph, Guelph, Ontario N1G 2W1, Canada}
\author{D.\,S.~Cross}
\affiliation{TRIUMF, 4004 Wesbrook Mall, Vancouver, British Columbia V6T 2A3, Canada}
\affiliation{Department of Chemistry, Simon Fraser University, Burnaby, British Columbia V5A 1S6, Canada}
\author{A.\,B.~Garnsworthy} 
\affiliation{TRIUMF, 4004 Wesbrook Mall, Vancouver, British Columbia V6T 2A3, Canada}
\author{R.~Kshetri} 
\affiliation{TRIUMF, 4004 Wesbrook Mall, Vancouver, British Columbia V6T 2A3, Canada}
\affiliation{Department of Chemistry, Simon Fraser University, Burnaby, British Columbia V5A 1S6, Canada}
\author{J.\,N.~Orce} 
\affiliation{TRIUMF, 4004 Wesbrook Mall, Vancouver, British Columbia V6T 2A3, Canada}
\affiliation{Department of Physics, University of the Western Cape, P/B X17, Bellville, ZA-7535, South Africa}
\author{M.\,R.~Pearson} 
\affiliation{TRIUMF, 4004 Wesbrook Mall, Vancouver, British Columbia V6T 2A3, Canada}
\author{E.\,R.~Tardiff} 
\affiliation{TRIUMF, 4004 Wesbrook Mall, Vancouver, British Columbia V6T 2A3, Canada}
\author{H.~Al-Falou} 
\affiliation{TRIUMF, 4004 Wesbrook Mall, Vancouver, British Columbia V6T 2A3, Canada}
\author{R.\,A.\,E.~Austin} 
\affiliation{Department of Astronomy \& Physics, Saint Mary's University, Halifax, Nova Scotia B3H 3C3, Canada}
\author{R.~Churchman} 
\affiliation{TRIUMF, 4004 Wesbrook Mall, Vancouver, British Columbia V6T 2A3, Canada}
\author{M.\,K.~Djongolov} 
\affiliation{TRIUMF, 4004 Wesbrook Mall, Vancouver, British Columbia V6T 2A3, Canada}
\author{R.~D'Entremont} 
\affiliation{Department of Astronomy \& Physics, Saint Mary's University, Halifax, Nova Scotia B3H 3C3, Canada}
\author{C.~Kierans} 
\affiliation{Physics Department, Simon Fraser University, Burnaby, British Columbia V5A 1S6, Canada}
\author{L.~Milovanovic} 
\affiliation{Department of Physics \& Astronomy, University of British Columbia, Vancouver, British Columbia V6T 1Z4, Canada}
\author{S.~O'Hagan} 
\affiliation{Department of Science, University of Alberta Augustana Campus, Camrose, Alberta T4V 2R3, Canada}
\author{S.~Reeve} 
\affiliation{Department of Astronomy \& Physics, Saint Mary's University, Halifax, Nova Scotia B3H 3C3, Canada}
\author{S.\,K.\,L.~Sjue} 
\affiliation{TRIUMF, 4004 Wesbrook Mall, Vancouver, British Columbia V6T 2A3, Canada}
\author{S.\,J.~Williams} 
\altaffiliation[Present address: ]{National Superconducting Cyclotron Laboratory, Michigan State University, East Lansing, Michigan 48824, USA.}
\affiliation{TRIUMF, 4004 Wesbrook Mall, Vancouver, British Columbia V6T 2A3, Canada}
\begin{abstract}
We report a precise determination of the $^{19}{\rm Ne}$ half-life to be $T_{1/2} = 17.262 \pm 0.007$~s. This result 
disagrees
with the most recent precision measurements and is important for placing bounds on predicted right-handed interactions that are absent in the current Standard Model. We are able to identify and disentangle two competing systematic effects that influence the accuracy of such measurements. Our findings prompt a reassessment of results from previous high-precision lifetime measurements that used similar equipment and methods.  
\end{abstract}
\date{\today}
\maketitle
Precise measurements of decay rates and angular correlations in semi-leptonic processes are known to be excellent probes for interactions that are predicted by extensions of the Standard Model~\cite{severijns:11}. For example, the measured lifetime and electron asymmetry in neutron $\beta$ decay~\cite{nico:09} are used to probe for right-handed currents and obtain a precise value of $V_{ud}$, the up-down element of the Cabibbo-Kobayashi-Maskawa quark-mixing matrix, in a relatively simple system that is free of nuclear structure effects. However, in spite of this compelling advantage, precision neutron $\beta$ decay experiments are challenging. Current results from independent neutron decay measurements show large discrepancies that need to be addressed before conclusive interpretations can be made from the data~\cite{nico:09}. 
In this regard Nature offers a fortuitous alternative in $^{19}{\rm Ne}$ $\beta$ decay, where the nuclear structure corrections are small and well understood~\cite{severijns:08}. Analogous to the neutron, the $\beta$ decay of the isospin $T = 1/2$ $^{19}{\rm Ne}$ nucleus proceeds 
predominantly via a {\it mirror} $1/2^+ \rightarrow 1/2^+$ transition to the analog ground state in $^{19}$F.  Furthermore, the vector and axial-vector form factors for the \iso{19}Ne case are similar to neutron decay. These similarities make $^{19}{\rm Ne}$ $\beta$ decay studies an attractive alternative to neutron decay experiments, without the experimental difficulties usually associated with the latter.
It has been shown that the $ft$ value (comparative half-life) for $^{19}$Ne decay, together with the measured $\beta$ asymmetry~\cite{calaprice:75},  allows a determination of $V_{ud}$~\cite{cuncic:09} and constitutes one of the most sensitive probes for right-handed interactions in the charged weak current~\cite{hostein:77,cuncic:91}. 
Presently the dominant contribution to the total uncertainty in the measured $ft$ value is from the $^{19}$Ne half-life.
Several measurements of the half-life have been reported that are in poor agreement with one another~\cite{severijns:08}. 

In this Letter we report a precise measurement of the \iso{19}Ne half-life, performed at the \triumf~Isotope Separator and Accelerator (\isac) facility in Vancouver, Canada.
In this experiment, radioactive $^{19}$Ne ions at $\sim$~37~keV were delivered to the 8$\pi$ gamma-ray spectrometer, an array of 20 symmetrically placed Compton-suppressed high-purity germanium detectors. The inner volume of the array comprised of 20 similarly placed 1.6-mm-thick BC404 plastic scintillator detectors called \sceptar~(Scintillating Electron Positron Tagging Array)~\cite{ball:05}, which were coupled to Hamamatsu H3695-10 photomultiplier tubes (PMTs) and covered $\sim$~80\% of the total solid angle. 
The ions were implanted on a $\sim$~1.3 cm wide, 40~$\mu$m-thick mylar-backed aluminium tape at the center of the spectrometer, which allowed us to register data using tape cycles.  
In each cycle we took background data for the initial 2 seconds, collected the \iso{19}Ne ions for $\sim$~1 second and reserved a counting time of 300 seconds~($\sim$~20~half-lives). At the end of each cycle 1~second was allocated for the tape movement to move any existing long-lived radioactivity away from the detectors, to a lead-shielded collection box placed downstream of the array center.
\begin{figure}[t]
\includegraphics[width=0.43\textwidth]{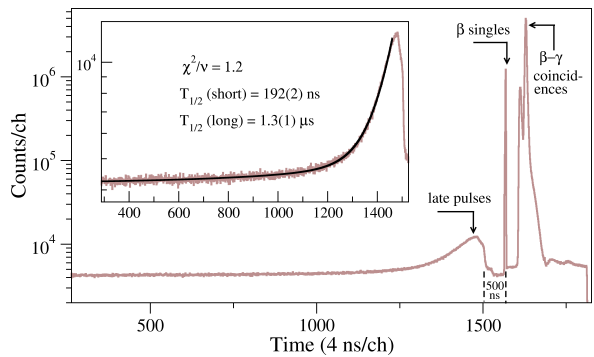}
\caption{(Color online) Summed TDC spectrum from all 20~\sceptar~detectors. The $\beta-\gamma$ coincidence, prescaled $\beta$ singles and the delayed prescaled $\beta$ singles events are identified. The CFDs were operated in ``blocking'' mode with a width of $\sim$500 ns. This prevented them from re-triggering within 500 ns of the first pulse. The inset shows a close up of the afterpulsing events from the scintillator detectors and a fit to the data.
}
\label{fig:tdc}
\end{figure}  
The $\beta$ signals from each \sceptar~PMT were shaped by a fast preamplifier and input to a constant fraction discriminator (CFD) whose output pulses were sent to a logic OR fan-in/fan-out unit. Eight outputs from the multi-channel fan-in/fan-out module were subjected to fixed non-extendible dead times in the range of 4-24~$\mu$s. This time range was much longer than any other signal time in the electronics (500~ns from the CFDs), or the response of the pulses from the PMTs, thereby ensuring that pulse pile-up would have negligible effect. The counts from the dead-time-affected streams were registered using a VME-based Struck SIS3801 multi-channel scaler (MCS) module. The time standard for the MCS data streams was provided by a 10~MHz~$\pm$~0.1~Hz temperature stabilized precision clock. Prescaled $\beta$ singles (with a prescale factor of 255) and $\beta-\gamma$ coincidence events were also recorded simultaneously in full list mode to obtain a measurement of the mirror ($1/2^+ \rightarrow 1/2^+$) branch. We 
defer discussion on this part of the analysis for a separate paper~\cite{triambak}. 

To test the ramifications of various systematic effects, we acquired the data at different initial rates and bin-widths, while the PMT bias voltages and CFD thresholds were kept unchanged. Two important systematic effects that influence the outcome of such high-precision lifetime measurements are detector afterpulsing and PMT gain shifts. Afterpulsing depends on the type of scintillator and PMT used and leads to the detection of spurious counts so that the deduced half-life would appear to be lower than the true value. The low-amplitude afterpulses mainly arise from long-lived phosphorescence in the scintillator material and the ionization of residual gas in the phototubes. On the other hand, PMT gain shift is a competing process that leads to event losses at high rates. These losses are caused by a fraction of the counts falling below the detection threshold and lead to an overestimation of the decay half-life. Below we describe the methods used to minimize the contributions of both these systematic effects 
and the procedure used to identify and disentangle them. 

To minimize the effects of afterpulsing, we raised the CFD thresholds while simultaneously viewing the pulses on an oscilloscope. This procedure reduced the contribution of the afterpulses, without discriminating against real events. However, a time-to-digital converter (TDC) spectrum obtained from the event-by-event data stream, shown in Fig.~\ref{fig:tdc}, revealed that a considerable number of the delayed pulses survived the raised CFD thresholds. The observed distribution of the remnant afterpulsing consisted of two exponentially decaying components with largely different decay constants. The areas under the prompt $\beta$ singles peak and the delayed afterpulses showed the afterpulsing probability to be~$\sim$~23\%.
\begin{figure}[t]
\includegraphics[width=0.43\textwidth]{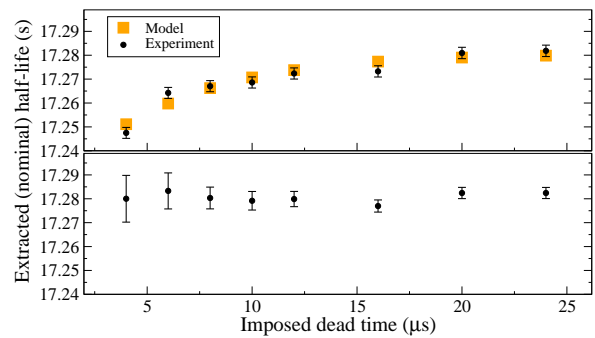}
\caption{(Color online) Extracted half-lives for a subset of the data (acquired at initial rates of $\sim$1 kHz per detector, with 1~s bin-widths) corrected for different dead times. The upper panel shows a significant systematic effect that is more pronounced at lower dead times, where the fits to the data yielded worse $\chi^2$. Normalized Monte Carlo simulation results are shown for comparison. 
The lower panel shows the same data set, corrected for afterpulsing using the results of our simulations. 
}
\label{fig:corrs}
\end{figure}
The first step in the data analysis was to preselect the data by fitting the individual cycles from each experimental run to a function that consisted of an exponential plus a flat background, using the maximum likelihood method described in Ref.~\cite{grinyer}. The few cycles that yielded anomalously deviant 
half-life values were tagged and subsequently excluded from further analysis. These comprised $\sim$~3\% of the total data used for the final analysis.
Data from the remaining `good cycles' were then summed and fit using the same procedure. 
For dead time corrections to these data, the imposed dead times were determined using the source+pulser technique~\cite{baerg}. The results from this analysis are shown in the upper panel of Fig.~\ref{fig:corrs}. We obtain discordant half-life values for the same data that were corrected for different dead times. This is the first time that such an effect has been observed and reported in a \mbox{scintillator-based,} high-precision half-life measurement. 

To better understand this systematic effect, we performed Monte Carlo simulations of the events using the summed TDC spectrum as a model. 
In the simulation, the relative intensities and time distribution of the afterpulses that follow $^{19}$Ne events were defined by the results of the fit shown in Fig.~\ref{fig:tdc}. 
Simulated events that occurred outside a user-defined dead time window were registered separately. These data were binned, dead-time-corrected, and fit using the same method described above. The fits to the simulated data showed a similar underestimation of the extracted half-lives at the lower dead times. 
\begin{figure}[t]
\includegraphics[width=0.425\textwidth]{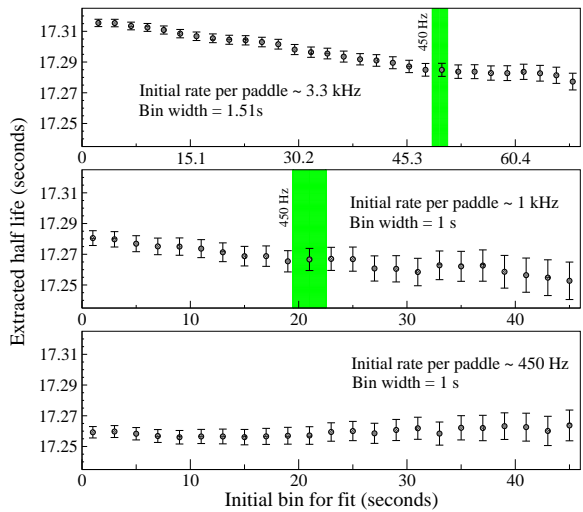}
\caption{(Color online) Half-lives obtained by successive fits to data acquired at different rates. 
The bottom panel shows results from the summed data of four runs (representing all the low rate data), 
while the other two panels show data from single runs. The bands in the high rate data show the time interval (after the beam was turned off) in which the rates would have dropped to $\sim 450$~Hz per detector.}
\label{fig:chops}
\end{figure}
The general agreement between our model and the data, shown in the upper panel of Fig.~\ref{fig:corrs}, allowed us to make a first-order correction for the deviations. This was done using the following procedure: First, the simulated histograms were fit with an additional exponential component in the model, so that the modified fit function was 
\begin{equation}\label{eq:func}
y_{fit} = \int_{t_l}^{t_h} a_1~e^{\frac{-\ell n 2}{a_2}t}~dt + \int_{t_l}^{t_h} a_1~a_3~e^{\frac{-\ell n 2}{a_4}t}~dt + a_5~,
\end{equation}
where $\Delta t = t_h - t_l$ defines the width of the time bin.  
In the fits, the $a_2$ parameter was kept fixed at the input value for the \iso{19}Ne half-life, while the other parameters were allowed to vary. 
The $\chi^2$ probabilities of these modified fits were found to be markedly better for the low dead-time data (ranging from 15\% to 50\%), with the fits converging to a value of $a_4= 8.81 \pm 0.11$~s at all dead times. 
The best-fit value for the $a_3$ parameter (which yields the relative intensities of the two components) depended on the initial rates input to the simulations and decreased as expected at larger dead times due to the presence of fewer afterpulses in the dead-time-corrected histograms. 
On fitting the experimental data similarly, with the $a_3$ and $a_4$ parameters kept fixed at the values determined by the simulations, we observe a restoration of the expected consistency ({\it c.f.}~lower~panel,~Fig.~\ref{fig:corrs}), which lends support to the validity of our simulations. 
Finally, additional simulations showed that the afterpulsing required small corrections to the fixed dead times determined 
using the prescription of Ref.~\cite{baerg}. These corrections (of the order $\lesssim 1\%$) were incorporated accordingly. 

While the {\it alignment} of the extracted half-lives at different dead times shown in Fig.~\ref{fig:corrs} validates our understanding of the afterpulsing, the accuracy of the absolute half-life is further influenced by other systematic effects that need to be taken into consideration. In the analysis described below, we investigate the other dominant systematic effect due to PMT gain shifts. All fits described henceforth incorporate the corrections due to afterpulsing, using the function described in Eq.~\eqref{eq:func}.
We tested for rate dependent losses due to gain shifts by acquiring data at different initial trigger rates. 
As shown in Fig.~\ref{fig:chops}, we observe significant losses at higher rates, which result in the extraction of an erroneously large half-life. As an additional check for rate-dependent shifts, we successively fit the histograms over a decreasing range of data points. In each fit, the initial time bin (that defined the first channel of the range over which the fit is performed) was increased in steps of \mbox{1-2}~channels, and the half-life value was recorded.
Fig.~\ref{fig:chops} shows the results from three scenarios, with data acquired at two extreme initial rates and an intermediate value of $\sim$~1~kHz per detector. We show only the results from the data that were subjected to the largest dead time of 24~$\mu$s, with minimal afterpulsing correction.  
 
Although the data showed no obvious signs of rate-dependent shifts at initial rates of $\sim$450~Hz, it was apparent that at the higher rates the PMT gains shifted significantly. Additionally, the data also showed that the phototubes needed a certain recovery time, depending on the intensity of the rates. This is obvious in the upper panel of Fig.~\ref{fig:chops}, where even after the detector rates had fallen to $450$~Hz, the rate-independence shown in the bottom panel is not reproduced.     
Mindful of these effects we used only a subset of the total data from the experiment for the final analysis. These only included data that were acquired at initial rates of $\sim$450~Hz and $\sim 1$~kHz per detector.  
For the data acquired at initial rates of $\sim$1~kHz, we waited for a conservative time period of $\sim$~two half-lives (34~seconds) after the beam was turned off to start fitting the data. 
With a relatively low background rate ($\sim 10~{\rm s}^{-1}$), this approach ensured that the accuracy of our result was not compromised due to the recovery time of the PMTs mentioned above. 

\begin{figure}[t]
\includegraphics[width=0.425\textwidth]{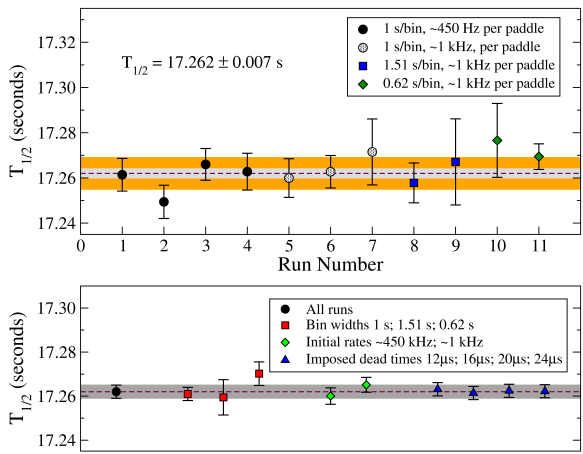}
\caption{(Color online) Top panel: Extracted half-lives for different runs at three different bin widths. The dotted line is the weighted mean, the lighter band shows the statistical uncertainty, and the darker band represents the final uncertainty in the half-life. Bottom panel: A comparision of the half-life values obtained under different conditions to the weighted mean from all runs (with statistical uncertainties only).  
}
\label{fig:results}
\end{figure}

Using the analysis described above we obtain the \iso{19}Ne half-life to be \mbox{$T_{1/2} = 17.262 \pm 0.007$~s.} The individual results from the runs are shown in the top panel of Fig.~\ref{fig:results}, where the central value in each run was obtained by taking the mean of the half-life values obtained at the higher dead times of 12-24~$\mu$s (where the corrections due to afterpulsing were the smallest). The contributions of the systematic and statistical uncertainties to our result are shown in Table~\ref{tab:errors}. The systematic uncertainty is dominated by gain shifts. We conservatively estimate the uncertainity due to gain shifts using the results obtained on varying the initial bin for the fits by $\pm 12$~seconds. 
Another systematic effect that could bias the result of this experiment is the potential long-lived diffusion of neon atoms from the tape. We assume this to be negligible over the time intervals in this experiment. Our assumption is based on previous measurements, where no apparent neon diffusion was observed using the same tape~\cite{grinyer:07,*grinyer_thesis}. The bottom panel in Fig.~\ref{fig:results} shows that our measurement is not significantly affected by other sources of systematic uncertainties.
\begin{table}[b]
\begin{flushleft}
\caption{Uncertainty budget for the half-life measurement.}
\label{tab:errors}
\begin{ruledtabular}
\begin{tabular}{lc}
Source & ${\rm Uncertainty (s)}$\\
\colrule
Afterpulsing corrections$^\dagger$ &$0.003$\\
Gain shifts &$0.006$\\
Dead time corrections &$0.0005$\\
Statistics & $0.002$\\
Total &$0.007$
\end{tabular}
\end{ruledtabular}
$^\dagger$ Obtained from our Monte Carlo simulations.
\end{flushleft}
\end{table}

\begin{figure}[t]
\includegraphics[width=0.425\textwidth]{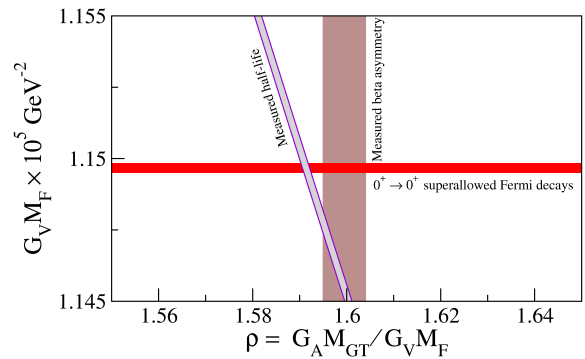}
\caption{(Color online) 
In the above, the bands represent the $\pm 68\%$ confidence intervals. $G_V = G_F V_{ud}g_V$, where $g_V$ is the vector form factor. In the low momentum transfer limit, $g_V(q^2 \rightarrow 0) = C_V = 1$. 
}
 \label{fig:bands}
 \end{figure}

Our result for the \iso{19}Ne half-life disagrees with the most recently published high-precision result, \mbox{$17.219 \pm 0.017$~s \cite{az:75},}  
by approximately 2.5 standard deviations.
Using the present value of the half-life, the $Q_{EC}$ value of the decay~\cite{geithner}, the analog branch~\cite{severijns:08} and the electron capture fraction~\cite{bamb,toi}, we obtain the corrected ${\cal F}t$ value of this mixed transition using the equation 
\begin{eqnarray}
\nonumber {\cal F} t & = &f_V t (1+\delta_R')(1 + \delta_{NS}^V - \delta_C^V) \\
& = &\frac{K}{G_V^2 M_F^2(1 + \Delta_R^V)(1 + \frac{f_A}{f_V}\rho^2)}
\end{eqnarray}
$= 1721.3 \pm 1.2~{\rm s}$. 
Here $G_V = G_F V_{ud} C_V$ is the effective vector coupling constant, and we follow the definitions and notation described in Refs.~\cite{severijns:08,cuncic:09}.
The resulting ${\cal F}t$ value is nearly a factor of two more precise than the previously adopted value~\cite{cuncic:09}. Together with the vector coupling constant, this value can be used to predict the beta asymmetry for \iso{19}Ne decay assuming only $V-A$ interactions and the conserved vector current hypothesis. 
We find the Standard Model beta asymmetry parameter to be \mbox{$A_\beta^{\rm S.M} = -0.0416 \pm  0.0007$}, which can be compared to the experimentally measured value, \mbox{$A_\beta^{\rm Exp} = -0.0391 \pm 0.0014$}~\cite{calaprice:75}.
This comparison is shown in terms of the mixing ratio and the vector coupling constant in Fig.~\ref{fig:bands}, where the constraints set by our half-life measurement can be seen. The data show a hint of disagreement which invites further investigation. 
A remeasurement of the beta asymmetry parameter (which dates to 1975) is welcomed at this point.   

In conclusion, our measurement provides useful data to complement ongoing precision neutron decay experiments and is particularly relevant for searches 
of right-handed interactions in the charged weak current. We observe a combination of systematic effects that need to be considered for an accurate result. 
We speculate whether similar effects, particularly detector afterpulsing, could have gone unnoticed in other high-precision half-life measurements that used plastic scintillators.

We thank I.\,S.~Towner for calculating the statistical rate function. We also benefited greatly from fruitful discussions with A.~Garc{\'i}a and O.~Naviliat-Cuncic.
This work was partially funded by a Natural Sciences and Engineering Research Council of Canada grant. \mbox{\triumf}\ receives federal funding via a contribution agreement through the National Research Council of Canada.
\bibliographystyle{apsrev4-1.bst}


\begin{thebibliography}{9}
\bibitem{severijns:11} N.~Severijns and O.~Naviliat-Cuncic, Ann. Rev. Nucl. Part. Sci. \textbf{61}, 23 (2011).
\bibitem{nico:09}J.\,S.~Nico, J. Phys. G: Nucl. Part. Phys., \textbf{36}, 104001 (2009).
\bibitem{severijns:08}N.~Severijns, M.~Tandecki, T.~Phalet, and I.\,S.~Towner, Phys. Rev. C \textbf{78}, 055501 (2008).
\bibitem{calaprice:75}F.\,P.~Calaprice \textit{et al.}, Phys. Rev. Lett \textbf{35}, 1566 (1975).
\bibitem{cuncic:09}O.~Naviliat-Cuncic and N.~Severijns, Phys. Rev. Lett \textbf{102}, 142302 (2009).
\bibitem{hostein:77}Barry R.~Holstein and S.\,B.~Treiman, Phys. Rev. D \textbf{16}, 2369 (1977).
\bibitem{cuncic:91}O.~Naviliat-Cuncic, T.\,A.~Girard, J.~Deutsch, and N.~Severijns, J. Phys. G: Nucl. Part. Phys. \textbf{17}, 919 (1991).
\bibitem{ball:05}G.\,C.~Ball \textit{et al.}, J. Phys. G: Nucl. Part. Phys. \textbf{31}, S1491 (2005).
\bibitem{triambak}S.~Triambak \textit{et al.}, in preparation.
\bibitem{grinyer}G.\,F.~Grinyer \textit{et al.}, Phys. Rev. C \textbf{71}, 044309 (2005).
\bibitem{baerg}A.\,P.~Baerg, Metrologia \textbf{1}, 131 (1963).
\bibitem{grinyer:07}G.\,F.~Grinyer \textit{et al.}, Phys. Rev. C \textbf{76}, 025503 (2007).
\bibitem{grinyer_thesis}G.\,F.~Grinyer, Ph.D. thesis, University of Guelph (2007).
\bibitem{az:75}G.~Azuelos and J.\,E.~Kitching, Phys. Rev. C \textbf{12}, 563 (1975).
\bibitem{geithner}W.~Geithner \textit{et al.}, Phys. Rev. Lett \textbf{101}, 252502 (2008).
\bibitem{bamb}W.~Bambynek \textit{et al.}, Rev. Mod. Phys. \textbf{49}, 77 (1977).
\bibitem{toi}R.\,B.~Firestone \textit{et al.}, eds., in Table of Isotopes (Wiley, New York, 1966) 8th ed.
\end{thebibliography}
\end{document}